\documentclass[sigconf]{acmart}

\setcopyright{rightsretained}
\acmDOI{}
\acmISBN{}
\acmConference[MCC]{Technical Report MCC.2019.1}{December 2019}{Berlin, Germany}
\acmBooktitle{Technical Report MCC.2019.1}
\acmYear{2020}
\copyrightyear{2020}

\usepackage[utf8]{inputenc}
\usepackage[english]{babel}
\usepackage[autostyle=true]{csquotes}
\usepackage{url}

\usepackage{tabularx}
\usepackage{multirow}
\usepackage{booktabs}
\usepackage{pdfpages}

\usepackage{amsmath,amssymb,amsfonts}
\usepackage{graphicx}
\usepackage{textcomp}
\usepackage{xcolor}

\usepackage{balance}

\usepackage{algorithm}
\usepackage{algpseudocode}

\usepackage{todonotes}
\begin{document}
\includepdf[pages=-, scale=1.0]{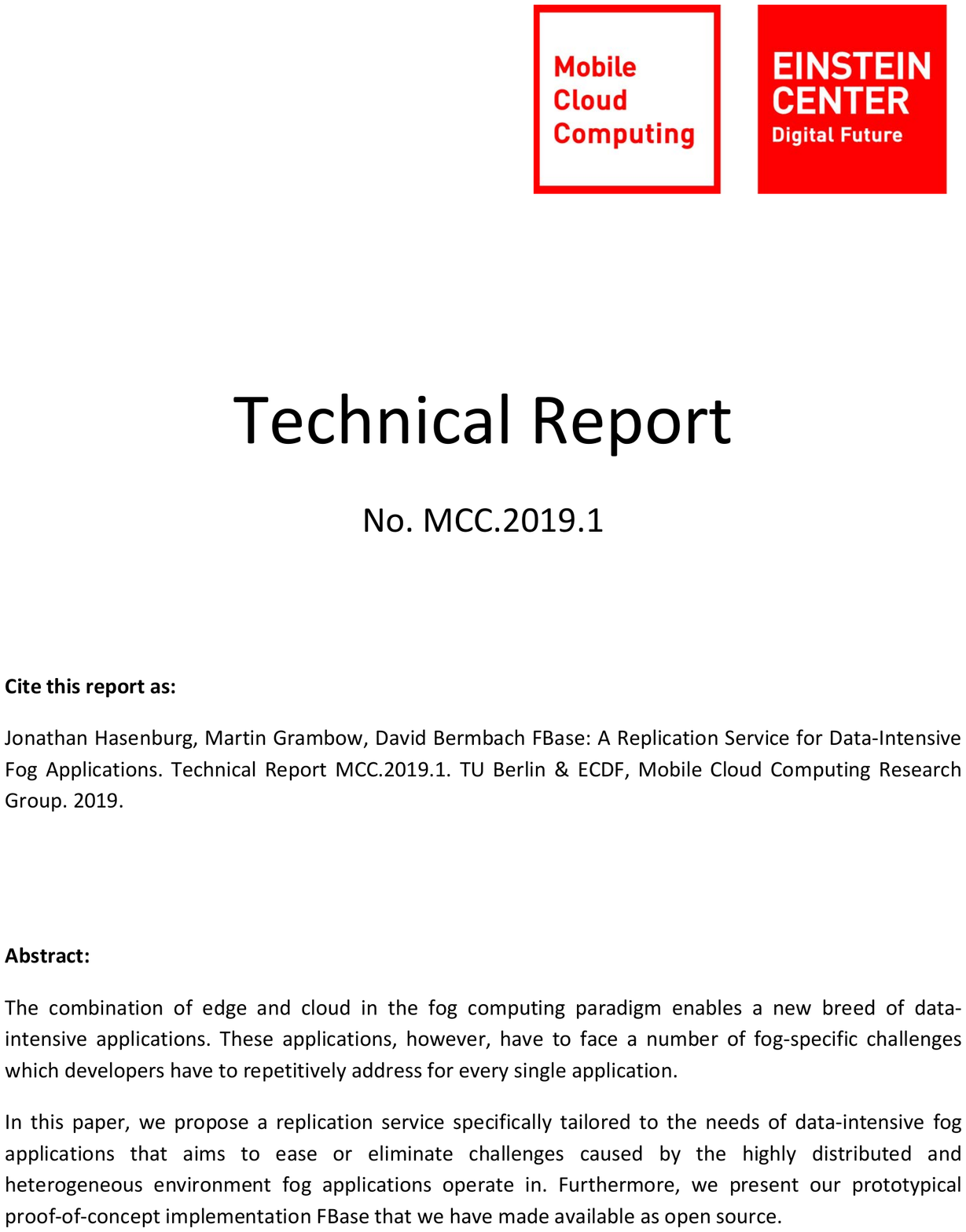}

\title{FBase: A Replication Service for Data-Intensive Fog Applications}

\author{Jonathan Hasenburg, Martin Grambow, David Bermbach}
\affiliation{
    \institution{TU Berlin \& Einstein Center Digital Future \\ Mobile Cloud Computing Research Group}
    \city{Berlin, Germany}
}
\email{{jh, mg, db}@mcc.tu-berlin.de}

% \author{Jonathan Hasenburg}
% \affiliation{
%     \institution{TU Berlin \& ECDF \\ Mobile Cloud Computing Research Group}
%     \city{Berlin, Germany}
% }
% \email{jh@mcc.tu-berlin.de}
%
% \author{Martin Grambow}
% \affiliation{
%     \institution{TU Berlin \& ECDF \\ Mobile Cloud Computing Research Group}
%     \city{Berlin, Germany}
% }
% \email{mg@mcc.tu-berlin.de}
%
% \author{David Bermbach}
% \affiliation{
%     \institution{TU Berlin \& ECDF \\ Mobile Cloud Computing Research Group}
%     \city{Berlin, Germany}
% }
% \email{db@mcc.tu-berlin.de}

\renewcommand{\shortauthors}{J. Hasenburg et al.}

\begin{abstract}
The combination of edge and cloud in the fog computing paradigm enables a new breed of data-intensive applications.
These applications, however, have to face a number of fog-specific challenges which developers have to repetitively address for every single application.

In this paper, we propose a replication service specifically tailored to the needs of data-intensive fog applications that aims to ease or eliminate challenges caused by the highly distributed and heterogeneous environment fog applications operate in.
Furthermore, we present our prototypical proof-of-concept implementation FBase that we have made available as open source.
\end{abstract}

% <concept_id>10002951.10002952</concept_id>
% <concept_desc>Information systems~Data management systems</concept_desc>
% <concept_id>10002951.10002952.10003400</concept_id>
% <concept_desc>Information systems~Middleware for databases</concept_desc>
% <concept_significance>100</concept_significance>
% <concept_id>10010520.10010521.10010537</concept_id>
% <concept_desc>Computer systems organization~Distributed architectures</concept_desc>
% <concept_significance>300</concept_significance>
% \ccsdesc[500]{Information systems~Data management systems}
% \ccsdesc[100]{Information systems~Middleware for databases}
% \ccsdesc[300]{Computer systems organization~Distributed architectures}


\maketitle

\section{Introduction\label{sec:intro}}
%!TEX root = ../main.tex

Current state-of-the-art applications are typically deployed on top of cloud services; the cloud alone, however, is often not capable enough for emerging application domains such as autonomous driving, 5G mobile applications, eHealth, or the Internet of Things (IoT)~\cite{paper_zhang_cloud_is_not_enough_GDP}.
Depending on the use case, reduced end-user latency, bandwidth limitations between sensors and cloud, or privacy challenges force developers to use fog computing instead: the combination of edge and cloud computing but also with optional small- to medium-sized data centers in the network between cloud and edge offers the best from both worlds~\cite{paper_bermbach_fog_vision,paper_shi_fog_computing_definition}.

Due to these benefits, we have seen a number of fog applications emerge over the last few years, e.g.,~\cite{paper_pradhan_chariot_edge_iot,paper_hou_vehicular_computing,paper_chen_fog_video_surveillance,paper_grambow_fog_video_surveillance}; however, one would still expect much more adoption of the fog computing paradigm.
In~\cite{paper_bermbach_fog_vision}, a number of possible reasons such as a lack of edge services or even hardware heterogeneity are discussed.
Beyond these, we also see the problem of having to ``reinvent the wheel'': developers need to start virtually from scratch for every fog application to \textit{get data to where it is needed by a multi-tenant application in a highly distributed and heterogeneous environment}.

In this paper, we propose a replication service which provides a set of middleware abstractions specifically tailored for this task.
These abstractions allow application developers to specify data placement and data movement using a declarative programming style while actual data distribution across multiple fog nodes is handled by the underlying replication service which we named FBase.
In essence, our approach is novel in that FBase provides the abstractions and infrastructure for application-controlled replica placement.
Therefore, we make the following contributions\footnote{This is an extended version of~\cite{paper_hasenburg_towards_fbase}.}:
\begin{itemize}
    \item We identify a set of requirements for a replication service that aims to simplify the development of data-intensive fog applications (section~\ref{sec:requirements}).
    \item We propose and describe our replication service FBase and discuss how it addresses the identified requirements (section~\ref{sec:design}).
    \item We present our proof-of-concept prototype, which we have made available as open source, and its evaluation (section~\ref{sec:eval}).
\end{itemize}
We also discuss related work (section~\ref{sec:relwork}) and our approach (section~\ref{sec:disc}) before drawing a conclusion (section~\ref{sec:concl}).

%When using the abstractions provided by FBase, developers of data-intensive fog applications no longer need to handle problems such as cluster membership management or access control.
%Instead, they provide configuration details to the middleware platform which hides the underlying complexity.

\section{Requirements\label{sec:requirements}}
%!TEX root = ../main.tex

Data-intensive fog applications encounter a number of challenges; most of them are not new but they are significantly more pronounced than in existing cloud-based systems and, thus, require new solutions.
The two most obvious challenges are the geo-dis\-tri\-bu\-tion and heterogeneity of the runtime infrastructure.
While cloud-based systems may run in a few geo-dis\-tri\-bu\-ted data centers on top of more or less identical VM hardware, a fog-based system runs on a variety of machines.
These can range from single board computers such as a Raspberry Pi to clusters of cloud VMs and anything in between, geo-distributed over at least hundreds of sites.
For data-intensive applications, this means to handle replication and data distribution in such an environment.

Beyond these, resources at or near the edge are limited so that fog-based systems need to deal with much higher degrees of shared resources, not only at the level of infrastructure resources but also in the software stacks on top of that.
Finally, fog-based systems need to interface with a variety of existing systems. These could be embedded cyber-physical systems at the edge, event brokers of all kinds, or stream processing systems and legacy applications in the cloud.

Ideally, a service as proposed in this paper deals with all these aspects to let application developers experience the same simplicity in the fog that they got used to while building cloud applications.
We believe that this ideal situation is not achievable, e.g., complete distribution transparency is not feasible in the presence of faults.
A well-designed service, however, should provide suitable abstractions to handle the complexities of, e.g., geo-distribution, while not hiding the fact per se.
Based on these premises, we identified the following main requirements for a replication service supporting data-intensive fog applications:

\textbf{Design for Multi-Tenancy:}
Due to the higher degree of shared resources near the edge, it is not feasible to run several instances of the replication service (or alternative services) in parallel.
Instead, such a service should be designed for multi-tenancy out of the box.

\textbf{Application-Controlled Data Placement:}
Applications should be able to declaratively specify the placement of data.
This means that they should control the placement of data while the underlying service handles data replication, movement, and distribution where and when necessary.
This still exposes the fact that different sites exist but takes the hassle out of it.

\textbf{Hiding Infrastructure Heterogeneity:}
An application should not have to worry about the number and kind of available machines at a particular site.
Instead, these should be exposed through suitable abstractions, e.g., the total amount of available resources, while the service handles load balancing, scheduling, and resource management.

\textbf{Ability to Interface with Existing Systems:}
A data-intensive fog application very likely needs to interact with other applications and systems, particularly at the edge but also in the cloud.
A data-handling service should provide this functionality as part of the data management tasks as data-intensive applications will interact with other systems to either ingest or expose data.
The application should be able to specify such interfaces in the same declarative way that it uses for data placement.

\section{FBase Design\label{sec:design}}
%!TEX root = ../main.tex

The main goal of FBase is to provide applications with the means to control data replication and data flow across geo-distributed sites using a declarative programming style.
As a typical example, consider the following application: an IoT sensor produces data at the edge (e.g., a temperature sensor), triggers a local actuator (e.g., a smart blind), and buffers data at a nearby edge node.
This data is then replicated to an intermediary node, e.g., at the regional level, where it is merged with data from other sensors, aggregated, and then replicated to the cloud.
In the cloud, the aggregated historical data from a multitude of sensors is made available to web-based clients.

In this example, it is irrelevant to the application which physical machine handles the data at each site.
As such, we developed the concept of nodes to \emph{hide infrastructure heterogeneity}.
In FBase, a node is a set of machines at a specific geographic location.
Nodes self-organize and application clients (or other nodes for that matter) can choose to interact with any machine of a given node.
We describe this concept in more detail in section~\ref{sec:nodes}.

In addition, the \emph{application can control data placement} and data flows in a declarative way.
For this, we propose the concept of keygroups which group data items that should be handled in the same way; each keygroup has metadata that describes which FBase nodes are handling the keygroup.
FBase nodes can be involved in two roles (simultaneously): first, as replica nodes which persist data locally and serve application requests (e.g., get or put); second, as trigger nodes which passively listen to any updates on the keygroup data and expose these updates via an event-based interface.
One purpose of the latter is to \emph{interface with existing systems}; in our example above, this means triggering the local actuator at the edge and triggering the data merge on the intermediary node.

Node details, the geo-location of sites, and other information are available to applications through our cloud-based naming service. We describe the naming service in more detail in section~\ref{sec:naming_service} and the details of keygroup and data distribution in sections~\ref{sec:keygroups} and~\ref{sec:distribution}.

Finally, since keygroups are entirely isolated from each other, FBase is inherently \emph{designed for multi-tenancy}.
We describe more details on this in section~\ref{sec:access}.

Beyond addressing the requirements from section~\ref{sec:requirements} as outlined above, we also describe the resulting consistency model of FBase in section~\ref{sec:consistency} and how we imagine building on top of FBase in future work in section~\ref{subsec:summary}.

\subsection{Naming and Configuration Management}
\label{sec:naming_service}

FBase offers a number of tuning knobs through which applications can, for instance, control how data is replicated.
Furthermore, for tasks such as access control or infrastructure management, it is necessary to assert that machines have unique IDs and that configuration data is stored with strict consistency guarantees.
In contrast, application data can often tolerate eventual consistency~\cite{paper_vogels_eventually_consistent}.

For configuration data, we decided to follow a similar design as in GFS~\cite{paper_ghemawat_google_file_system} and BigTable~\cite{paper_chang_bigtable} which both use Chubby~\cite{paper_burrows_chubby} to handle configuration data in a consistent way.
Our ``Chubby'' is a component called \emph{naming service} which handles naming (i.e., assignment of unique IDs) and storage of configuration data.
IDs are immutable and are tombstoned when they are no longer needed.
This means that other machines can safely cache configuration data and also share it with other machines so that the naming service itself is only involved when adding or removing machines.
Overall, the naming service acts as the single point of truth for configuration data and naming in case of conflicts.

As the naming service stores no application data, which is likely to be updated frequently, its load is usually low; hence, it is unlikely to become a scalability bottleneck.
Furthermore, it can easily be sharded to mitigate potential scalability problems.

We explicitly decided to introduce the naming service component instead of handling all naming and configuration management in a peer-to-peer (P2P) fashion for four reasons.
First, having a consistent view on configuration data significantly simplifies the design and implementation of the rest of FBase.
Second, especially when moving towards the edge, network bandwidth and hardware resources may be very limited and unreliable.
Therefore, it is preferable to handle these tasks only on machines with sufficient resources.
Having a dedicated component allows us to select these machines as part of the deployment process instead of having to reconfigure FBase instances to either run or not run this functionality.
Third, a dedicated naming service can be scaled independently of the rest of the system as needed.
Fourth, the naming service itself is only a component and does not make assumptions about its own implementation.
In fact, our prototype comes with two implementations.
Both are essentially wrappers that expose their internal strictly consistent storage structure as the naming service interface.
One is based on the local file system and the other on Apache Zookeeper\footnote{zookeeper.apache.org}; additional ones can be added easily.

Even though we believe that these reasons speak for our approach, in certain situations a \enquote{P2P naming service} can be a better option as, for example, configuration data cannot be updated when the naming service is unavailable. However, our remaining design aims to tolerate such situations temporarily, as explained in the following.

\subsection{Nodes: Managing Collocated Machines}
\label{sec:nodes}

\begin{figure}[ht]
\centering
\includegraphics[width=\columnwidth]{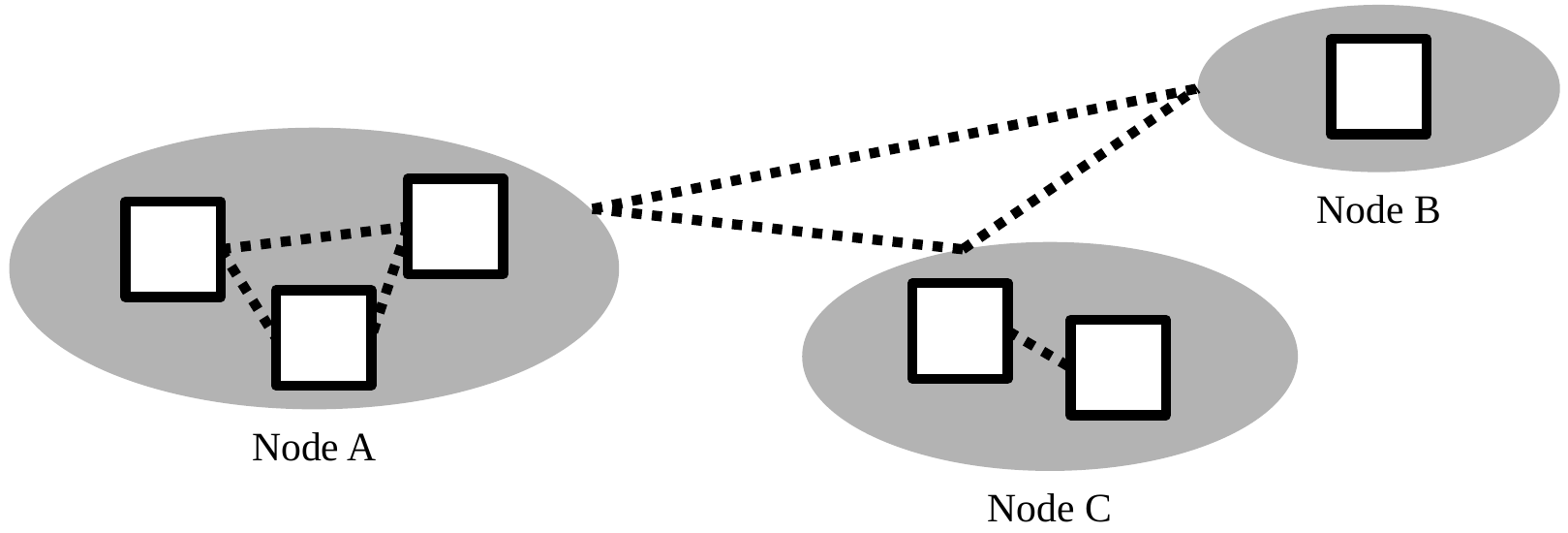}
\caption{FBase is a P2P System of Nodes which Comprise a P2P System of Machines Each}
\label{fig:global_node_cluster}
\end{figure}

To hide the complexity caused by infrastructure heterogeneity and geo-distribution, FBase provides an abstraction for collocated machines called \emph{nodes}, as such machines often have the same purpose and are used to scale-out systems.
For instance, in an IoT scenario where sensor data is preprocessed at the edge before sending it to a cloud backend, a machine at the edge has to handle data of only a very limited number of sensors while the cloud backend has to handle the data from all edge devices.
An obvious solution to this is to have several (preferably stateless) cloud machines behind a load balancer with a shared storage system.
Therefore, nodes are groups of collocated machines at the same geographical site that have a shared purpose and use a shared data storage system for persistence.
Overall, this means that FBase has two levels of infrastructure abstraction: on a node level, FBase is (more or less) a P2P\footnote{As the naming service handles some coordination logic, it is not a P2P system in the purest sense.} system of nodes (see also figure~\ref{fig:global_node_cluster}); within a node, FBase is a P2P system of machines with a shared persistence tier.

This allows us to strictly separate data distribution functionality from scaling mechanisms: nodes have a unique name so that messages and data are always addressed to a node instead of to a specific machine.
The machine of the target node that ends up processing this message is hidden from all entities outside of that target node\footnote{Communication in FBase is done via pub/sub; nodes internally distribute responsibility for other nodes and the responsible machine subscribes to all updates from the sender node's machines.}.
The strict separation of responsibilities is also helpful for infrastructure membership management: at the node level, there will only be very low churn and even temporary unavailabilities of nodes can be expected to be infrequent due to the built-in redundancy.
Also, this helps to keep the load on the naming service at a low level as the more frequent machine churn can be handled within nodes without involving the naming service.

\begin{figure}[ht]
\centering
\includegraphics[width=\columnwidth]{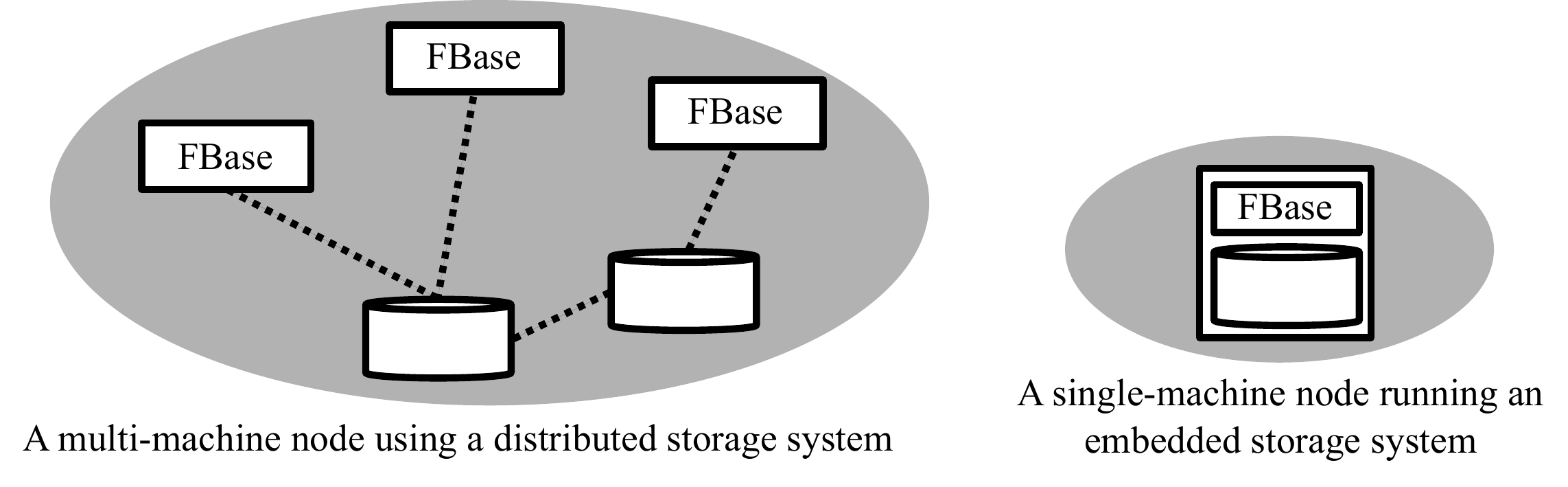}
\caption{FBase Nodes can be Single Machines with Embedded Storage or Clusters of Machines with Shared Storage Systems}
\label{fig:node_database}
\end{figure}

In its simplest form, e.g., at the edge, a node is only a single machine running FBase and either an embedded database system or the local file system as a ``shared'' persistence tier.
On the other end, e.g., in the cloud, a node may comprise a cluster of machines running FBase and either a dedicated shared storage system or a shared part (e.g., a database table) of a cloud storage service.
In this design, we assume that the connectivity of nodes may be poor while connectivity within a node is mostly fast and stable.
Figure~\ref{fig:node_database} shows two example setups, one medium-sized node with a dedicated shared storage tier and one small node, e.g., a single Raspberry Pi, using the local file system for persistence.
For production deployments, we generally recommend collocating instances of the storage system and FBase to facilitate low latency data access.

What happens at node level is based on configuration data stored in the naming service (see details in section~\ref{sec:distribution}).
Within a node, however, machines must be able to self-manage, as they might not be able to connect to other nodes or some central server.
This self-management comprises management of cluster membership, fail-over in case of machine crashes, load distribution, etc.
We use the shared storage system for ``communication'' within a node so that machines can remain stateless.
Furthermore, storing configuration data along with application data has the additional benefit that machines that can serve application requests also have access to the latest configuration data as well.

\subsection{Keygroups: Encapsulating Logically Coherent Data}
\label{sec:keygroups}
Applications often have groups of data items that should be handled in the same way, i.e., they should use the same access policies, should be replicated in the same way, and will often be queried together.
Typical examples for this are a sequence of time series values produced by a specific sensor or user records that would be stored in the same table of a relational database.
Comparable to the entity groups in Megastore~\cite{paper_baker_megastore}, FBase uses the concept of so-called \emph{keygroups} for handling such groups of logically coherent data items which allows natural sharding.

Each keygroup has a globally unique name, some keygroup metadata, as well as the actual data records.
The keygroup metadata holds a number of application-defined policies which specify how data should be replicated (see section~\ref{sec:distribution}) and who should have access to the data (see section~\ref{sec:access}).
This is the key mechanism to give applications control over data distribution.

While keygroup creation and metadata updates (such as giving another party access) need to be confirmed by the naming service, each involved node also stores a copy to reduce load on the naming service and to reduce latency.
This naming service dependence ensures that keygroups have globally unique names and that only authorized parties can access the respective data records.

The data records within a keygroup each comprise an unordered set of key-value pairs along with some per-record metadata, in particular update timestamps.
As this abstraction is (on purpose) very similar to the BigTable~\cite{paper_chang_bigtable} interface, it is relatively simple to utilize any of the widely used, scalable column stores as node persistence tier.

For multi-tenant setups, we propose to use a keygroup naming scheme that maps a keygroup to its tenant, e.g., by including the application id, user id, and data record description.
This would lead to names such as ``SmartHomeApp.SomeUser.Temperatures''.

\subsection{Distribution of Data}
\label{sec:distribution}

FBase provides two primary mechanisms for data distribution that are both specified on a per-keygroup level in the keygroup metadata: replication and transmission.
For replication, a set of so-called \emph{replica nodes} is defined that (i) each stores a copy of the respective keygroup's data records and (ii) accepts updates on data records of that keygroup which are then forwarded to all other nodes that are part of the keygroup's node set.
Transmission, in contrast, is a mono-directional update propagation mechanism where the so-called \emph{trigger nodes} receive updates from the replica nodes but have only read access to the data records.
Trigger nodes specifically exist to integrate legacy applications or external systems such as a stream processing system through an event-based interface that exposes an update stream.
As defined in section~\ref{sec:requirements}, developers should not have to worry about infrastructure management and heterogeneity.
Thus, to replicate data to nodes in a specific geographic area, developers can query the naming service about adequate target nodes as the geo-location is part of the node metadata.
Then, the only action left is to add these nodes to the keygroup; FBase takes care of the rest.

\begin{figure}[t]
\centering
\includegraphics[width=\columnwidth]{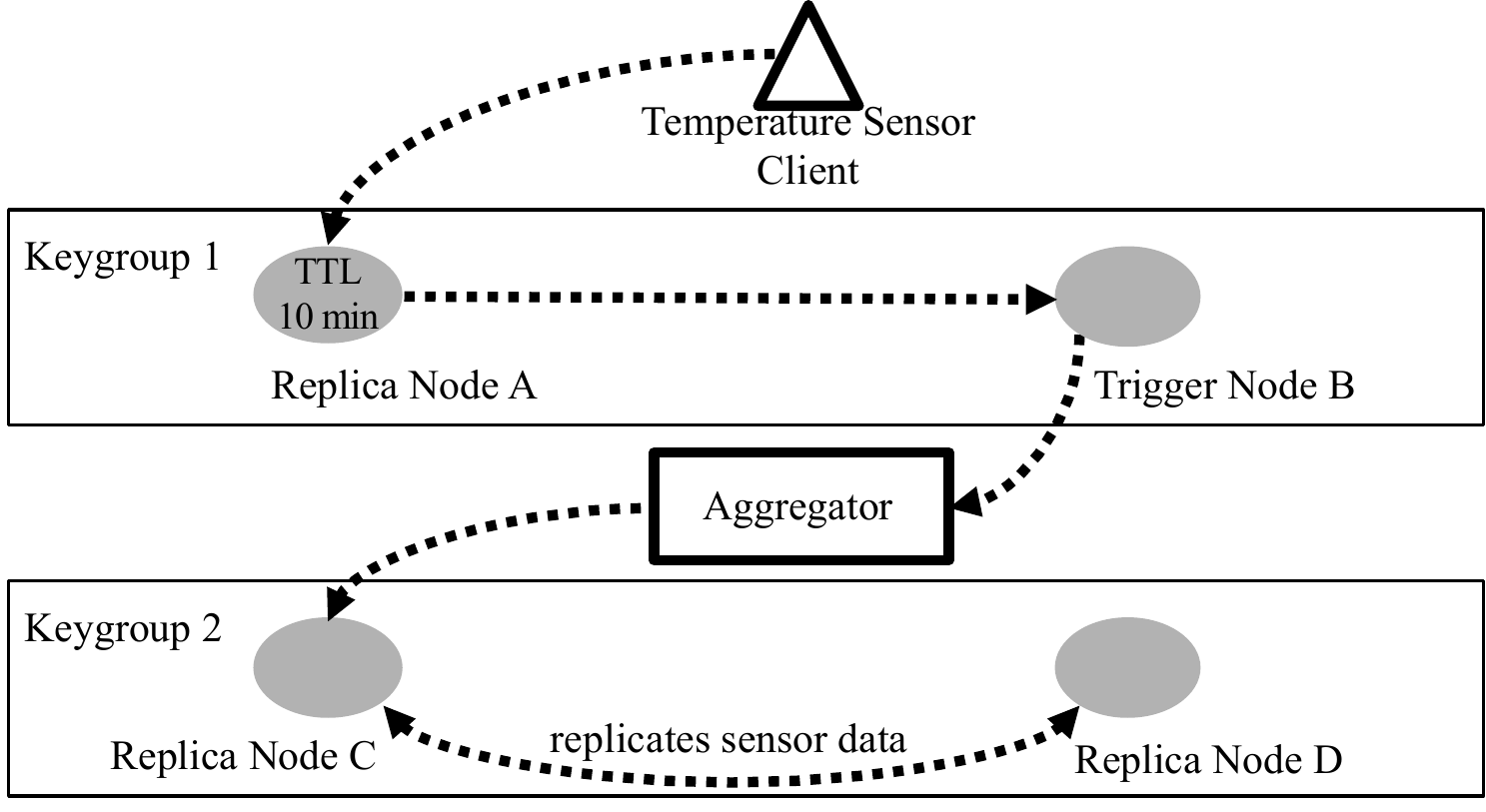}
\caption{Example: Using Replica and Trigger Nodes to Control Data Distribution}
\label{fig:distribution_control}
\end{figure}

In practice, replica nodes may store incomplete replicas as applications can also specify a time-to-live (TTL) for each replica node of a keygroup, i.e., these replicas will store data records only for a limited period of time.
This is particularly useful for edge nodes with limited storage capacity but can also be used to instantiate a buffered data stream that is accessible both as update stream and in an OLTP fashion.
The combination of node types and TTL allows applications to specify arbitrary data distribution schemes as needed.
See figure~\ref{fig:distribution_control} for a typical IoT example use case: temperature information is ingested at the edge and buffered there for 10 minutes as defined by Keygroup 1.
In addition, an external aggregator component based on a trigger node continuously reads the temperature information from the keygroup and forwards aggregated values to the cloud where they are stored persistently.
Keygroup 2 ensures that all the aggregated data is replicated to a second cloud node.
Note, that an FBase node can take the role of a replica node and trigger node of the same keygroup at the same time to fulfill both functionalities; in the example, a single node can act as Replica Node A and Trigger Node B.

For the actual distribution of data, FBase uses a publish-subscribe approach.
This supports loose coupling of communicating nodes and, thus, improves overall availability as nodes simply publish updates in a fire-and-forget way similar to update propagation in PNUTS~\cite{paper_cooper_pnuts}.
It is the responsibility of recipients to create the required subscriptions (based on the specifications made in the keygroup metadata).
While this decouples individual nodes -- which is most times the only feasible option for such geo-distributed deployments, it can lead to lost messages and out-of-order delivery.
To mitigate these message delivery problems, messages use a counter as ID and nodes buffer the keys of data records for which they sent their most recent update messages.
This way, recipients can detect missed updates and contact a sender node directly to request the latest version of the respective data record.
In most cases, this ensures that messages are delivered reliably, and data becomes eventually consistent.
However, a combination of machine failures, high update rates, and unfavorable buffer size and TTL settings may lead to permanently lost updates from the perspective of the receiving node.
This is then the price to pay for high availability, low latency, scalability, and wide range of supported data distribution configurations which are more important for many applications than small amounts of lost data (e.g., temperature values from a sensor).
Still, data loss can be avoided by leaving TTL disabled.

\subsection{Security: Tenant Isolation and Access Control}
\label{sec:access}
While security is a complex topic, there is a variety of state of the art solutions that should be used.
Furthermore, as FBase is about the programming abstractions for application developers, most questions of security are beyond the scope of this paper.
We would like, however, to briefly outline how we realize tenant isolation and (to a certain degree) access control in FBase.
For this, we need to protect both application data and configuration data from unauthorized access.

For the protection of application data, we assume that the network within a node is secure and that the machines involved are trustworthy.
For instance, machines can add themselves to a node if they can access its storage system.

However, when transmitting data records between nodes or to application clients, these records are encrypted using AES128 with a keygroup-specific secret that is stored along with the keygroup meta-data.
Instead of restricting access to data records, we only control access to this secret which is automatically updated whenever a node is removed from the keygroup's node set.

The communication necessary for the operation of FBase such as configuration changes is secured using RSA2048 (coupled with AES128), i.e., the data is encrypted with the recipients public key and signed so that the recipient can verify the identity of the sender.
Furthermore, we added some basic authorization policies:
\begin{enumerate}
    \item When starting FBase, an initial node and an application client are created. Only existing nodes and application clients can authenticate themselves with the naming service, which is necessary for creating additional nodes and application clients.
    \item Nodes cannot add themselves to existing keygroups, but they can add other nodes to their own keygroups.
\end{enumerate}

We believe that these measures provide sufficient security and tenant isolation under our assumed threat model.
FBase can, of course, be complemented with other state-of-the-art security mechanisms and technologies.

\subsection{Consistency Model\label{sec:consistency}}
For all of the following explanations, it is important to understand that we divide data into two groups: configuration data and application data.
Configuration data is stored with strict consistency guarantees as it describes the information required for the operation of FBase, e.g., information on cluster membership or keygroup secrets.
Application data, however, is distributed based on asynchronous pub/sub communication and is stored in multiple, often only eventually consistent, data stores.
Since nodes in geo-distributed fog networks are likely to fail and will often have high inter-node latency, we do not see a realistic alternative to asynchronous replication. Also, replication across eventually consistent data stores by itself would already result in eventual consistency~\cite{paper_bermbach_metastorage}.
Of the two consistency dimensions ordering and staleness, ordering is arguably the more challenging one for applications~\cite{paper_vogels_eventually_consistent,diss_bermbach}, and we cannot avoid staleness.
For ordering, however, we propose to follow the intuition behind the adaptive mastership in PNUTS~\cite{paper_cooper_pnuts} which is based on the observation that almost all data has a single writer, i.e., to size keygroups in a way that there will be only one concurrently writing client.
For multi-writer keygroups, we plan to add an append-only keygroup type in the future which is inspired by the append-only logs of~\cite{paper_zhang_cloud_is_not_enough_GDP}.

Beyond these, FBase does not provide any restrictions on application developers regarding the data distribution policies.
As such, developers can design distribution schemes that result in frequent data loss, e.g., by setting TTL to zero.
We do not want to restrict the configuration space in this regard as there might be applications that need precisely these configuration options, e.g., when only the latest data values are relevant.

\subsection{Summary and Vision\label{subsec:summary}}
FBase uses the concepts of nodes and keygroups to satisfy most of the requirements from section~\ref{sec:requirements}.

With nodes, FBase mostly hides the complexity of geo-dis\-tri\-bu\-tion and infrastructure heterogeneity. From an application perspective, it is irrelevant how many machines a node comprises, each group of one or more machines at the same site is simply assigned a unique node ID and self-manages.

With keygroups, applications can control the placement of data to define arbitrarily complex replication schemes (including preprocessing in between) using a declarative programming style.
In a similar fashion, applications can also integrate external systems such as legacy applications or stream processing systems.
Furthermore, keygroups make it straightforward to use shared resources in multi-tenant setups, as each tenant can have its own namespace.

FBase is also a first step towards a comprehensive fog data management system.
Specifically, we plan to work on predictive replica placement in which a prediction component will automatically adjust keygroup membership based on actual and anticipated physical client movement to ease the burden on applications while letting them retain control on replica placement.

\section{Evaluation\label{sec:eval}}
%!TEX root = ../main.tex

We evaluate FBase in three different ways: First, we give an overview of our proof-of-concept implementation that we created to show the feasibility of the design (section~\ref{sec:poc}).
Second, we present the results of several micro-benchmark experiments which we ran to evaluate the overheads created by FBase (section~\ref{sec:micro-bench}).
Third, we describe how FBase could be used for the implementation of fog application scenarios (section~\ref{sec:usage_scenarios}).

\subsection{Proof of Concept Implementation}\label{sec:poc}

To demonstrate the feasibility of our system design, we have implemented it as a proof-of-concept prototype in Java 8.
Our prototype consists of three separate components -- the FBase daemon, the naming service, and an application-side client library.
In this section, we give an overview of our prototype which is also available as open source\footnote{\url{https://github.com/OpenFogStack/FBase}}.

\paragraph{FBase Daemon}

The FBase daemon is the software component that runs on every individual machine.
Multiple machines running instances of the FBase daemon will organize themselves as a node if they are located at the same site and can connect to the same storage system.
Each daemon serves application requests, publishes incoming updates to subscribers, and stores updates in the node storage system which makes the data available to all other deamons running on machines of the same node.
The deamons also manage the subscriptions of the node, monitor the availability of other machines, and periodically verify whether the locally cached configuration information is still consistent with the state of the naming service.
When a daemon realizes that it missed some updates from another node, e.g., when a machine within its node crashed or in case of network disruptions, it requests re-transmission of the corresponding data records.

Daemons map to the node concept in the following way: within a node, machines distribute responsibilities via the shared storage system.
A machine that is responsible for handling keygroup ``my.key.group'' will subscribe to all machines of the other nodes in that keygroup for updates on the topic ``my.key.group''.
The list of machines that are part of a node is currently stored by the naming service so that external machines will notice changes in membership (and thus potentially missing subscriptions) via the regular interaction with the naming service; in that case, missed update requests will be detected and requested again upon the next update.
For future versions, we plan to only keep a set of seed machines listed in the naming service so that requests for node membership, i.e., other machines, are directed to the responsible node directly.

The FBase daemon uses an adapter architecture for connecting to storage systems; currently, it implements connectors for Amazon S3\footnote{aws.amazon.com/s3} (for cloud nodes) and in-memory storage (for single machine nodes).
To add more connectors, e.g., a Cassandra\footnote{https://cassandra.apache.org/} connector for small to medium sized nodes running in the fog, it is sufficient to implement our connector interface.
All communication between FBase nodes as well as between nodes and the naming service is based on ZeroMQ~\cite{zeromq} which offers additional message delivery guarantees beyond the ones implemented in FBase.

\paragraph{Naming Service}

The naming service fulfills the tasks described in section~\ref{sec:naming_service}; as mentioned, we currently have two implementations of it.
The first uses Apache ZooKeeper for consistent and replicated storage of configuration data.
In its current state, the implementation still requires some additional testing and bug fixing before it is ready to be used in practice.
Therefore, we also provide a second naming service implementation that mocks the use of ZooKeeper based on the local file system -- this could in fact be a viable distributed alternative when run on top of a distributed consistent file system.
This implementation can be run as a single machine naming service and we used it in our experiment runs described in section~\ref{sec:micro-bench}.

\paragraph{Application Client Library}

Applications can interact with FBase via a REST interface that is offered by every machine running the FBase daemon; we also implemented application-side client libraries in Java and Kotlin.
These client libraries expose the FBase interface in a more accessible way for all JVM-based applications.

\subsection{Micro-Benchmarks} \label{sec:micro-bench}

In this section, we present the results of some micro-benchmarks with which we want to demonstrate that our prototype works properly and that the performance and staleness overheads created by FBase are appropriate for a research prototype.

\subsubsection{Latency Overheads} \label{sec:latency_overheads}

With this experiment, we want to quantify the latency overhead for read and write operations, i.e., the latency difference between writing to a storage system directly or via FBase.
For this purpose, we added debug level logging to FBase and deployed it on an Amazon EC2\footnote{aws.amazon.com/ec2} m3.medium instance running the S3 connector\footnote{Obviously, a fog deployment would use other storage systems but our purpose here is only to evaluate the overhead of FBase.}.
On another m3.medium instance in the same availability zone, we installed a benchmarking client that sequentially issued 1000 put, read, and delete operations each to FBase. We measured both the end-to-end latency of FBase as well as the latency of writing from FBase to S3; table~\ref{tab:s3_stats} shows our results.

As one can see, the latency overhead of FBase (columns ``FBase'' minus ``S3'') for the read operations is rather small (10ms on average).
In this time, FBase receives a request via HTTP from another machine, parses the request, and returns the requested data record.
The overhead for write operations is larger, on average 87ms for put and 72ms for delete operations, since the updated data record also has to be provided to ZeroMQ which encrypts the data and publishes it asynchronously to other nodes (we publish data for this micro-benchmark even though there are no active subscribers deployed to evaluate the impact on the latency).
We include the overhead of publishing in these measurements since an operation does only commit when a data update has been published to increase durability (if there is a network partitioning and FBase crashes after applying an update, ZeroMQ will still deliver the message upon reconnection unless the machine crashes.

Overall, these numbers show that the FBase approach is feasible for non-realtime use cases.
It should be noted that our prototype has \emph{not} been optimized for performance in any way as our goal was only to demonstrate the feasibility of the programming abstractions provided by FBase and to give a ``proof of life'' of our prototype.

\begin{table}[t]
\centering
\caption{Latency Measurements in ms}
\label{tab:s3_stats}
\begin{tabular}{@{}lrrrrrr@{}}
\toprule
           & \multicolumn{2}{c}{Put}                             & \multicolumn{2}{c}{Read}                            & \multicolumn{2}{c}{Delete}                          \\
           & \multicolumn{1}{c}{FBase} & \multicolumn{1}{c}{S3} & \multicolumn{1}{c}{FBase} & \multicolumn{1}{c}{S3} & \multicolumn{1}{c}{FBase} & \multicolumn{1}{c}{S3} \\ \midrule
Min        & 50                         & 11                     & 10                         & 6                      & 40                         & 6                      \\
Max        & 1275                       & 1099                   & 990                        & 985                    & 1292                       & 1249                   \\
Avg        & 118                        & 31                     & 26                         & 16                     & 88                         & 16                     \\
Std Dev    & 97                         & 44                     & 45                         & 44                     & 92                         & 48                     \\
$Q_{0.95}$ & 215                        & 80                     & 65                         & 41                     & 185                        & 59                     \\
$Q_{0.99}$ & 540                        & 154                    & 138                        & 126                    & 410                        & 120                    \\ \bottomrule
\end{tabular}
\end{table}

\subsubsection{Staleness Overheads}
FBase distributes data asynchronously which means that replicas will incur staleness.
In this micro-bench\-mark experiment, we aimed to quantify the staleness introduced by FBase alone (not including response times of storage systems or network latencies) as an additional measure of FBase' compute overhead.
To achieve this, we deployed two FBase nodes and the benchmarking client on the same virtual machine; each running in their own Java VM.
For this purpose, we used an m3.xlarge instance which offers four virtual CPU cores so that our nodes and benchmarking client would not compete for compute resources.
We configured both FBase nodes to use the in-memory storage connector.

During the experiment run, the benchmarking client issued 1000 put and delete operations each as these two operation types can create temporary inconsistencies in FBase.
For each operation of the benchmarking client, we measured the time between completing the write on the first and on the second node, i.e., the data-centric staleness~\cite{diss_bermbach}; see table~\ref{tbl:staleness} for our results.
As one can see, the staleness overhead introduced by FBase itself is relatively low for both write operations.
Compared to network latency in a geo-distributed deployment (which FBase is designed for), these are negligible.

\begin{table}[t]
\centering
\caption{Staleness Measurements in ms}
\label{tbl:staleness}
\begin{tabular}{@{}lrr@{}}
\toprule
           & \multicolumn{1}{c}{Staleness FBase Put}    & \multicolumn{1}{c}{Staleness FBase Delete}    \\ \midrule
Min        & 6                          & 3                             \\
Max        & 98                         & 57                            \\
Avg        & 15                         & 8                             \\
Std Dev    & 11                         & 9                             \\
$Q_{0.95}$ & 36                         & 28                            \\
$Q_{0.99}$ & 60                         & 53                            \\ \bottomrule
\end{tabular}
\end{table}

\subsection{Possible Usage Scenarios with FBase}
\label{sec:usage_scenarios}
The main purpose of FBase is to simplify the development of data-intensive fog applications.
In this section, we describe three different usage scenarios; for each scenario, the implementation of an application would be difficult and labor intensive.
Using FBase, however, only the application logic needs to be implemented, as all data management tasks are handled by FBase based on configuration details, i.e., in all three scenarios the data handling code requires only put and get requests as well as a few configuration instructions.\footnote{We were originally planning to actually implement an example application but decided against it when we realized that almost all implementation effort would go towards implementing (completely unrelated) application-specific code.
Instead, we decided to show a code listing for the complete data handling code of the most complex use case (the mobile app scenario).}

\subsubsection{Remote Research Station}

\begin{figure}[ht]
\centering
\includegraphics[width=\columnwidth]{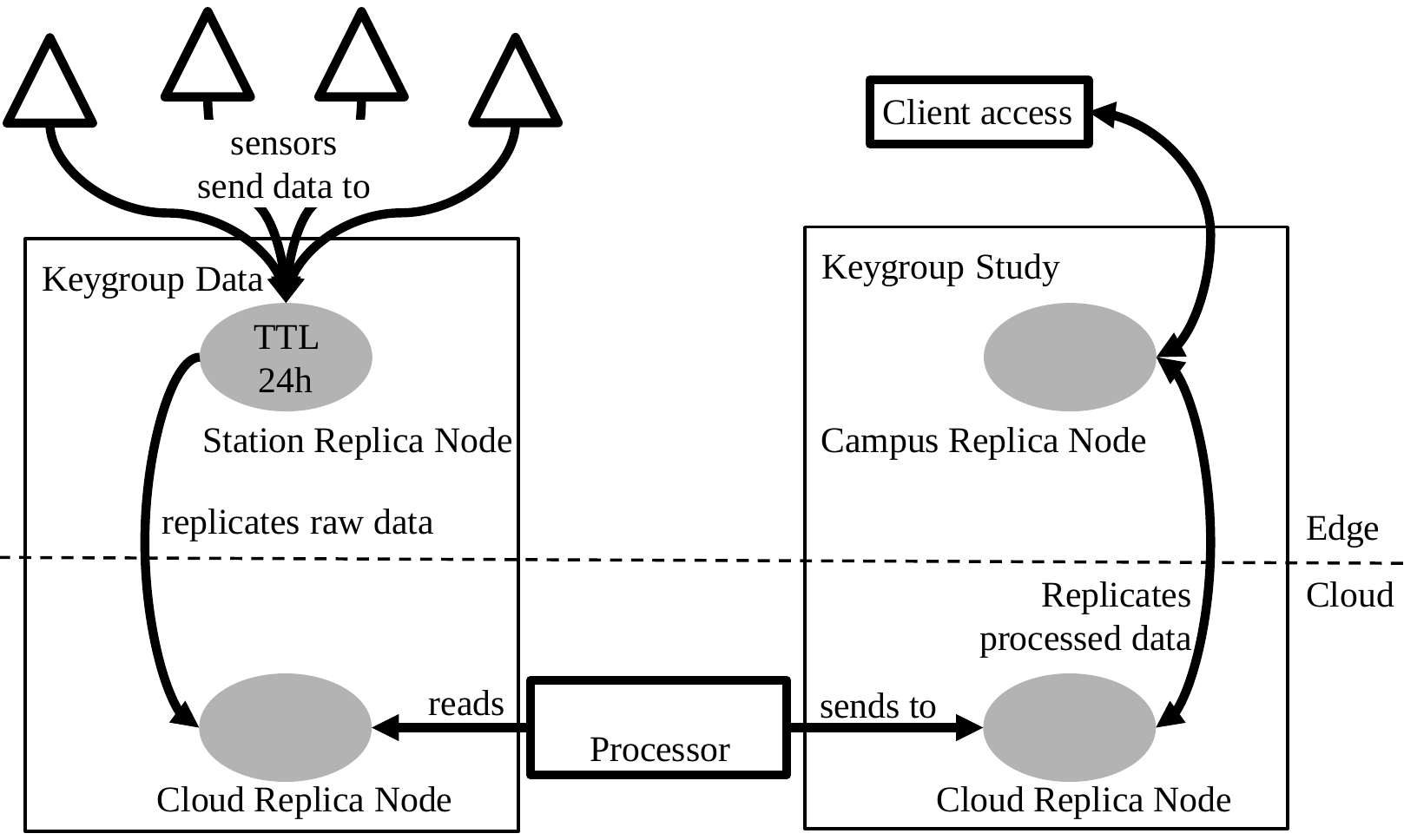}
\caption{Remote Research Station Scenario}
\label{fig:scenario_polar}
\end{figure}

Figure \ref{fig:scenario_polar} illustrates a scenario with an offsite research station, e.g., in the polar regions.
Here, recorded measurement information is buffered for one day at an edge node within the station and also replicated to a persistent cloud storage.
A cloud-based processor retrieves the data from the cloud node, analyzes it, and makes the preprocessed data available to universities and research centers worldwide.
For this, universities can evaluate the processed data on a local node which is automatically updated with all newly processed data.
In the figure, there could be additional edge nodes on other research campuses that are also part of the replica node set of the keygroup on the right.

% Required Code:
% Sensors:
%     data = get_data_from_sensor()
%     put(data)
%
% Processor:
%     data = read_new_data()
%     result = process(data)
%     put(result)
%
% Student:
%     data = read_specific_data()
%     update = do_research(data)
%     put(update)

\subsubsection{Carsharing Fleet Management}

\begin{figure}[ht]
\centering
\includegraphics[width=\columnwidth]{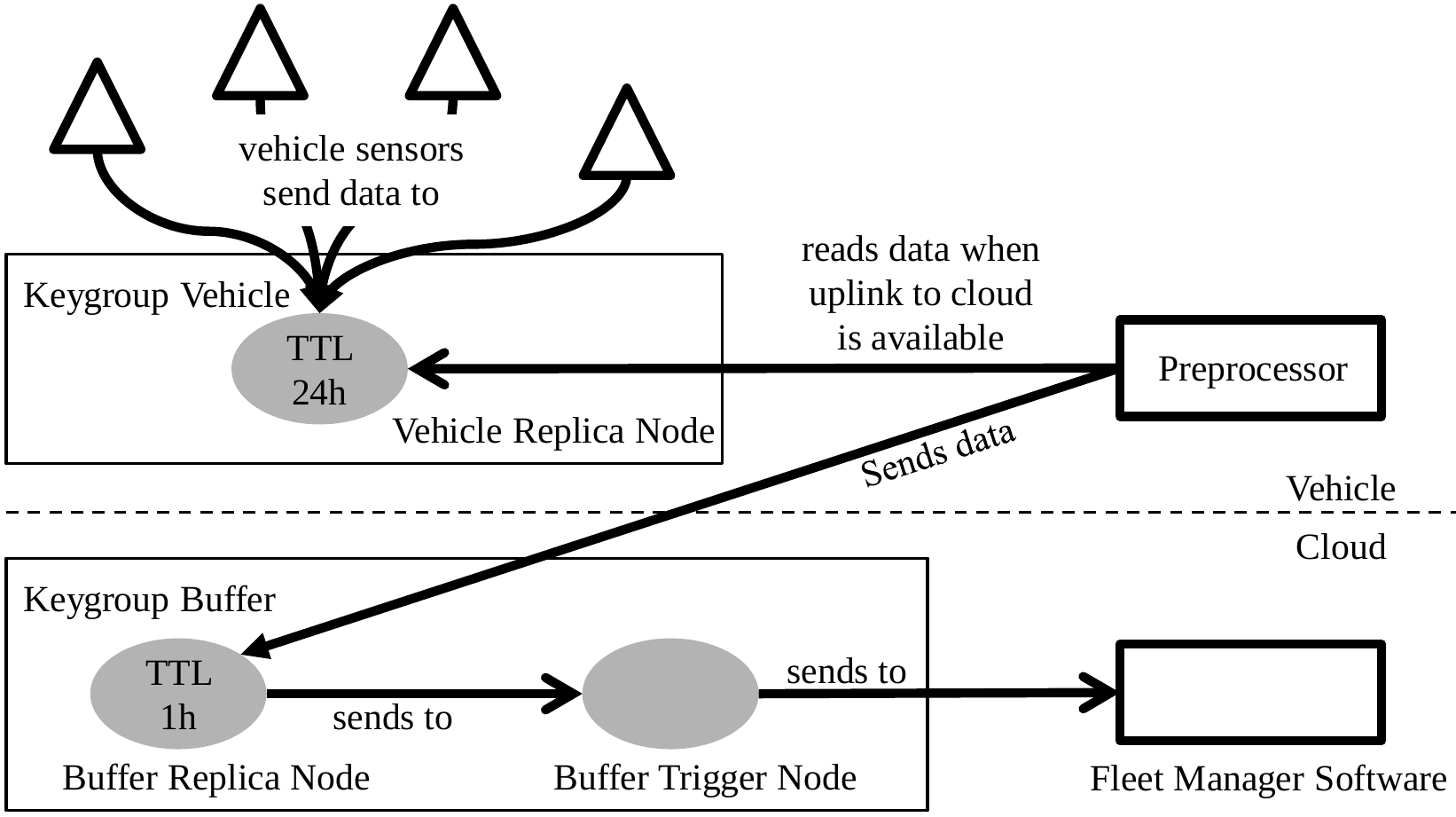}
\caption{Carsharing Fleet Management Scenario}
\label{fig:scenario_fleet_management}
\end{figure}

In the scenario shown in figure~\ref{fig:scenario_fleet_management}, vehicle sensors collect information on the maintenance state of the car so that the carsharing provider can schedule predictive maintenance along with the corresponding fleet capacity planning.
These sensors store their data inside the corresponding vehicle for 24 hours as defined by the related keygroup.
In addition, a preprocessor application, also located in the vehicle, retrieves the stored data whenever a good communication uplink to the cloud is available.
Then, it preprocesses the data, e.g., to reduce the size, and uploads the results to the Buffer Replica Node.
The keygroup of the uploaded data specifies that the data should be transmitted to a trigger node which here acts as an event-based connector to the fleet management software.
Similarly to figure~\ref{fig:distribution_control}, the same physical node can be both a replica node and trigger node, but it is also possible to use separate nodes.
The setup can easily support additional vehicles with the preprocessors of these vehicles also sending the data to the same Buffer Replica Node.

% Required Code:
% Sensors:
%     data = get_data_from_sensor()
%     put(data)
%
% Preprocessor:
%     wait_for_uplink()
%     data = read_new_data()
%     pre_processed = do_preprocessing(data)
%     put(pre_processed)
%
% Fleet Manager Software:
%     data = read_latest()
%     visualize(data)

\subsubsection{Mobile App with Moving Client}

\begin{figure}[t]
\centering
\includegraphics[width=\columnwidth]{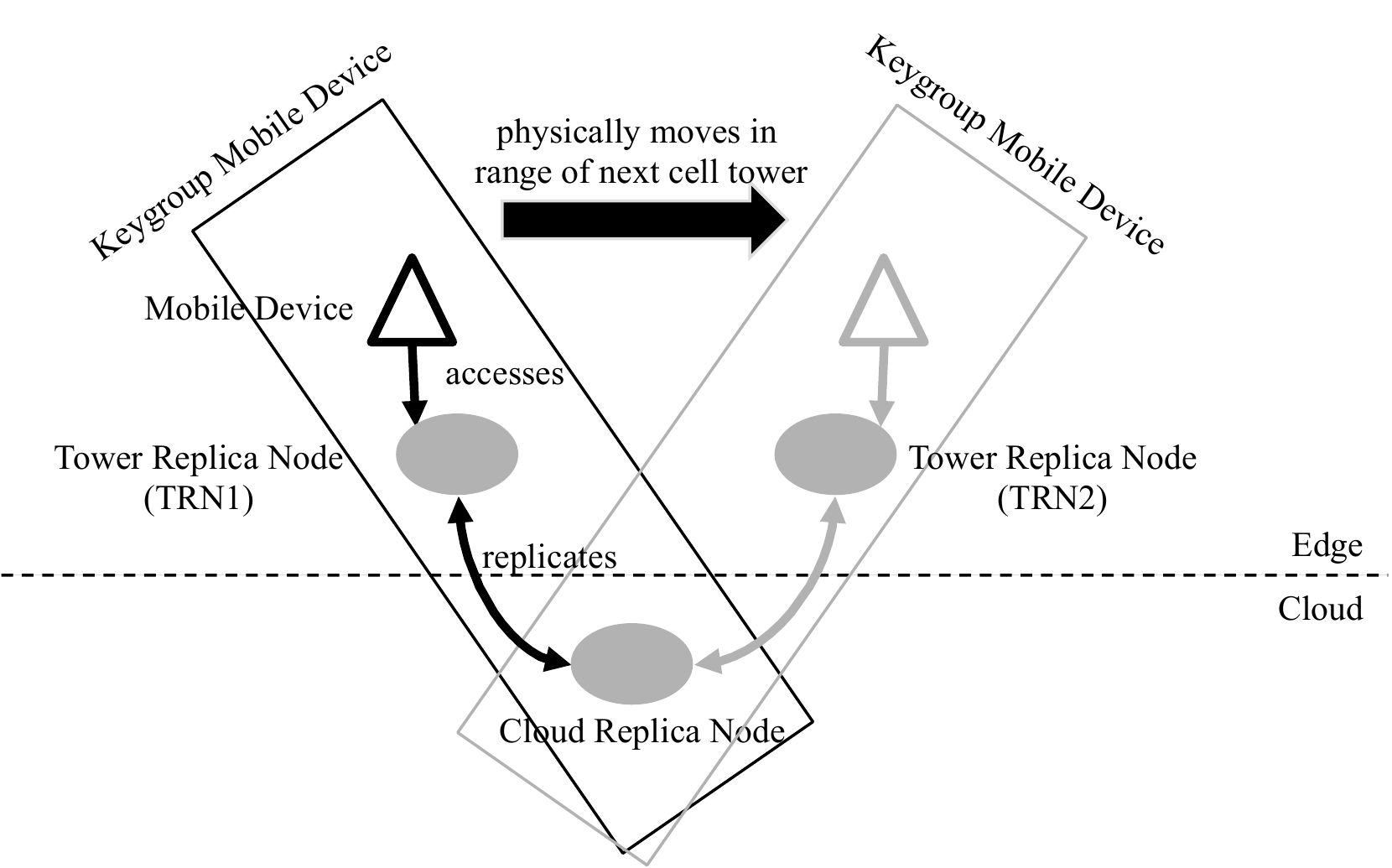}
\caption{Mobile App Scenario with Moving Client}
\label{fig:scenario_mobile_clients}
\end{figure}

In the scenario shown in figure~\ref{fig:scenario_mobile_clients}, a mobile device such as a smartphone or tablet runs an app that uses the cloud for persistence; to speed up data retrieval, data is also stored at the edge.
Depending on movement of the user, the respectively closest edge node should be used.
In the figure, keygroups are used to migrate data from one storage system -- here placed at a cell tower -- to another depending on the physical location of the client.
In the example, the user moves from tower 1 to tower 2 which initiates the update to the keygroup configuration: the replica node of tower 1 is replaced with the node at tower 2.
FBase can then either migrate the data between the two edge nodes or download it from the Cloud Replica Node to the Tower Replica Node 2 depending on latency and available bandwidth.

\begin{figure}[ht]
\centering
\includegraphics[width=\columnwidth]{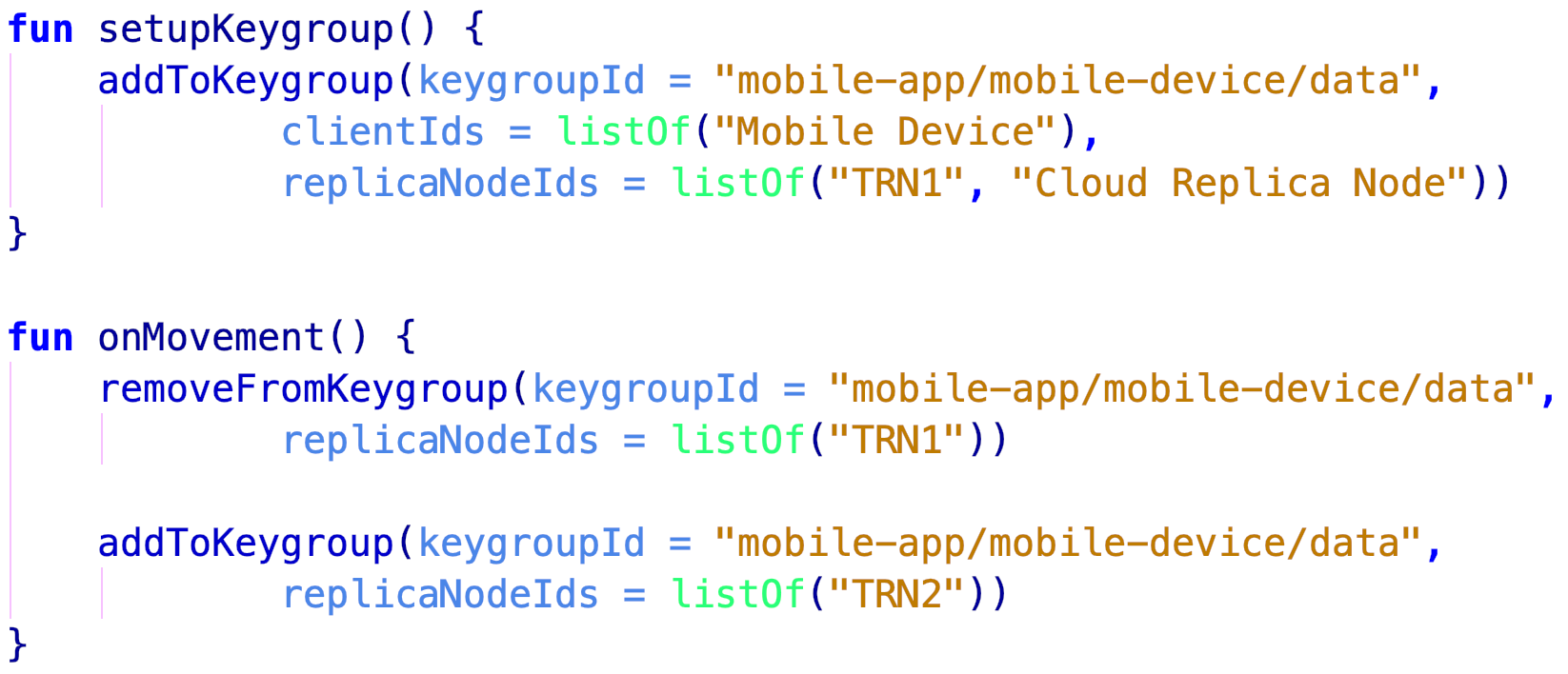}
\caption{Keygroup Setup and Client Movement Code}
\label{fig:scenario_mobile_clients_code}
\end{figure}

By just describing the desired results rather than required steps, it is effortless and straightforward to implement the replication and data management task with FBase.
In fact, most keygroup-related tasks can be accomplished with a single line of code, e.g., adding clients and replica nodes\footnote{Trigger nodes and time to live configurations can be added to the keygroup in a similar fashion, i.e., by adding additional parameters to the method.}.
Thus, the clients that operate in the fog environment can focus on their application-specific tasks.
These, however, can still be quite complicated, so we refrained from actually implementing a full-fledged fog application for this evaluation; the code needed to interact with FBase would comprise only a few lines while most effort would have to go into implementing application-specific logic.
See figure~\ref{fig:scenario_mobile_clients_code} which shows a Kotlin implementation of the FBase handling code using our Kotlin client library: the function \textsc{setupKeygroup()} should be invoked initially and (gets or) creates the described keygroup with two replica nodes.
The function \textsc{onMovement()} should be invoked whenever the mobile client connects to another cell tower.
In a real application, parameters like the replica node IDs would probably be provided as function parameters.

% Required Code:
% Sensors:
%     data = get_data_from_sensor()
%     if (movement) {
%         add_replica_node("Tower Replica Node 2")
%         remove_replica_node("Tower Replica Node 1")
%     }
%     put(data)

\section{Related Work\label{sec:relwork}}
%!TEX root = ../main.tex

Fog computing is still a relatively new computing paradigm.
As such, there are many open research questions left, e.g.,~\cite{paper_bermbach_fog_vision,paper_shi_fog_computing_definition,paper_diaz_challenges_iot_and_cloud,paper_shi_edge_challenges}, and there is only a limited number of existing publications in this research area.

Hourglass by Shneidman et al.~\cite{hourglass} is closely related to FBase as it targets a similar problem.
In Hourglass, so-called circuits describe the flow of IoT data from a data source to an end consumer.
In contrast to FBase, Hourglass only supports buffering and storage if it is explicitly defined and not by default.
Additionally, Hourglass does not consider replication of data items.
Overall, the circuit idea focuses on the end-to-end data stream abstraction while FBase uses keygroups and their fine-grained data distribution configuration which enables arbitrary data replication and distribution schemes.
This means that the basic unit in Hourglass is a single end-to-end stream where data is modified in between while FBase focuses on the individual stages where data is identical which can then be assembled into a larger data stream if required.
As such, FBase provides much more flexibility and is better suited to the goal of supporting data-intensive fog applications.
Furthermore, the focus on identical data sets instead of individual end-to-end streams avoids sending identical data over the same network connection several times while two circuits will always transfer the data twice.

With the Global Data Plane (GDP), Zhang et al.~\cite{paper_zhang_cloud_is_not_enough_GDP} propose a data management system design for IoT use cases.
GDP is built around the notion of single writer logs with append-only semantics that are broken down into chunks.
These chunks are placed inside a distributed hash table to support location-independent routing.
GDP seems to be in a very early development stage and does neither state specifics on replica placement nor offer an implementation.
In addition, it is bound to suffer from the downsides of hashing-based replica placement in the fog while FBase allows applications to freely specify arbitrary replication schemes as needed.
The data log abstraction itself is a good fit for IoT use cases but not for other fog application domains for which FBase' keygroup abstraction offers more flexibility.
In these applications, the GDP will quickly become overloaded as data can never be deleted so that it would benefit from a TTL feature as in FBase.

Mortazavi et al.~\cite{mortazavi2017cloudpath} describe CloudPath, a platform that migrates application logic and data among nodes along the path between edge and cloud.
These nodes can comprise one or multiple machines and use Cassandra for storage.
In difference to the data replication of FBase, the CloudPath data replication is indirectly controlled by application functions which are also deployed on these nodes.
CloudPath creates a local copy when the function needs access and removes the copy when the function has no further use of it.
FBase, on the other hand, lets the application directly control replication which is especially useful when replicas should be created preemptively or for other purposes than low latency data access, e.g., to increase data durability.
Furthermore, FBase supports applications running in any kind of execution environment while CloudPath applications have to run as platform-specific ``PathExecute'' containers.

In~\cite{paper_plebani_ditas_vision}, Plebani et al. describe their vision of so-called virtual data containers to support the development of data-intensive fog applications.
This approach seems to be in a vision stage and the focus is more on matching of application requirements to well-defined data sources controlled by other entities while FBase provides the building blocks for letting applications control data movement and replication of their own data.

With Nebula, Ryden at al.~\cite{paper_ryden_nebula} propose a grid-inspired distributed edge store.
Nebula centrally controls data placement in the so-called dataStore master.
In a large-scale geo-distributed deployment, however, directing all requests first to such a centralized master server is prohibitive in terms of performance, scalability, and availability.
In contrast, the naming service of FBase can be distributed and only (passively) handles metadata management rather than moving application data.

Based on their previous experiments~\cite{paper_confais_benchmarking_storage_for_fog}, Confais et al.~\cite{confais_object_2017} try to make IPFS fog-ready.
For this, they modify IPFS so that it does not read from the local file system but rather uses a local (per site) NAS system.
While this solves part of the data locality problem identified in their previous work~\cite{paper_confais_benchmarking_storage_for_fog}, IPFS is still based on the concept of immutable files which makes it more suitable for content delivery network use cases than for applications with frequent data updates: in such scenarios, IPFS will quickly hit the scalability ceiling.

To achieve lower latency for applications, Lin et. al \cite{paper_lin_enhancing_edge_computing} propose to locate a copy of the utilized database and an additional replication middleware also at edge nodes.
While all transactions are executed locally, the middleware tracks the changes and propagates the ``writesets'' to the central ``sequencer'' which propagates updates to all other edge nodes.
In contrast to FBase, this requires the sequencer as central coordinator which introduces a single point of failure.
Moreover, the sequencer relies on stable connections to the edge nodes whereas FBase is specifically designed for unreliable environments.
Finally, the sequencer uses full replication and does not allow arbitrary replication schemes as we do in FBase.

Psaras et al.~\cite{psaras2018mobile} also identify the need for data storage at the edge to reduce the stress on networking resoureses.
As a solution, they propose to use small data stores close to edge devices and discuss data management strategies which also include mobility aspects.
However, their paper is more about the implications for the business models of service providers while we propose and evaluate a technical solution.

While the papers discussed above primarily deal with data management and migration in fog environments, a number of papers specifically aim to identify the \enquote{optimal} node for data placement.
To solve this fog-related problem in geographically distributed systems with many instances, Gupta et al.~\cite{gupta2018datafog,gupta2018fogstore} as well as Mayer et al.~\cite{paper_mayer_fogstore} propose mechanisms to find suitable replication targets considering the physical data location and node stress levels.
Naas et al.~\cite{naas2017ifogstor} formalize this assignment problem and propose a heuristic approach for solving it.
For their approach, however, they require a lot of information on data characteristics and node capabilities which will often be hard to acquire, incomplete, and outdated in geo-distributed systems.
Nevertheless, combining such approaches with FBase' keygroups might be a promising avenue to pursue.

\section{Discussion\label{sec:disc}}
%!TEX root = ../main.tex

FBase provides powerful abstractions to data-intensive fog applications: instead of handling data distribution themselves, such applications can simply provide configuration details to FBase which then manages all data accordingly.
%We believe that this abstraction is nearly as powerful as the move from imperative programming to declarative programming for querying of database systems.
Using the abstractions of FBase, applications can create arbitrary complex data distribution setups with minimal efforts.
In fact, we planned to let students participating in one of our teaching projects implement the usage scenarios presented in section~\ref{sec:usage_scenarios} with the help of FBase, but had to learn that each scenario's implementation involves virtually no data handling code thanks to FBase.
However, for very simple use cases such as connecting a single sensor to a backend service, application developers might be better off implementing data management themselves.

At this point, we would like to emphasize again that FBase is \emph{not} a database system, FBase only offers the necessary infrastructure that allows developers to define replication paths, data flow, and fine-grained access control.
However, this infrastructure could easily be used to implement a distributed fog datastore -- in fact, we plan to do just that in future work: comparable to BigTable~\cite{paper_chang_bigtable} offering added functionality on top of GFS~\cite{paper_ghemawat_google_file_system}, we imagine a fog datastore running on top of FBase that automatically handles replica placement (preferably through predictive replica placement), replica selection, and infrastructure management based on application-side monitoring.

For our initial FBase design, we decided to use a BigTable-like~\cite{paper_chang_bigtable} data interface which is suitable for many but not all use cases.
As such, FBase should not be used to handle large, unstructured, binary data.
Likewise, applications that require more complex query functionality as provided by relational database systems or applications that primarily use timeseries data are not directly supported by FBase.
However, the overall design of FBase could easily be adapted to provide also relational keygroups or append-only logs for timeseries data.
Adding support for large unstructured binary data should be considered carefully: while it could easily be added as another keygroup abstraction, moving large data blobs around the fog may overload resources in practice; hence, the existing CRUD interface can be too coarse-grained for updates and data may need to be handled differently, i.e., not as a blob.

As already discussed in section~\ref{sec:distribution}, there are some unfavorable combinations of failure situations, TTL settings, replication configurations, and workload patterns which may not only lead to message loss but to actual data loss.
As a simple example, a keygroup comprising an edge and a cloud node each may use the edge node only for buffering before transmitting the data to the cloud node which then is intended for persistent storage.
With a TTL=60s setting for the edge node, any network connectivity problem between edge and cloud node that lasts longer than a minute will lead to data loss.
FBase does not protect developers from making such a design decision -- costs and benefits, particularly importance of data, should be carefully weighed.
For instance, choosing a larger TTL value or using an additional keygroup only comprising the cloud node and explicitly deleting all received data in the shared keygroup might be a better alternative than the presented example in case of important data.
FBase here provides several tuning knobs for the management of this tradeoff so that application developers can decide on a case by case basis how to handle it.

We have designed FBase for fog environments.
As a ``fog subset'', FBase can also be used to manage data distribution in a cloud-only deployment.
For example, it might manage data replication in a variety of cloud federation scenarios or it could be used to replace Amazon S3's cross-region replication with a more powerful alternative.

\section{Conclusion\label{sec:concl}}
%!TEX root = ../main.tex

Emerging application domains such as autonomous driving or the Internet of Things often rely on edge computing, e.g., to circumvent bandwidth limitations or to profit from low latency communication with end users.
However, as edge resources are inherently limited, fog computing as a paradigm which combines edge and cloud has recently emerged to combine the benefits of both worlds.
While there are already a number of fog applications, we see -- among others -- the lack of supporting tools and services for running on and integrating both edge and cloud as one of the main reasons for the slow adoption of the fog computing paradigm.

In this paper, we identified a set of requirements that a replication service for data-intensive fog applications should satisfy.
Based on these requirements, we proposed and described FBase, our replication service for data-intensive applications.
FBase, for example, allows applications to simply describe how data should be distributed rather than handling data management themselves.
As an evaluation, we presented our proof-of-concept prototype, the results of a number of micro-benchmark experiments, and three fog application scenarios that highly benefit from a service such as FBase.
\balance

\section*{Acknowledgements}
\noindent This work was partially funded by the Deutsche Forschungsgemeinschaft (DFG, German Research Foundation) - 415899119.

\bibliographystyle{ACM-Reference-Format}
\bibliography{bibfile}


\begin{thebibliography}{31}

%%% the Web address.

\ifx \showCODEN    \undefined \def \showCODEN     #1{\unskip}     \fi
\ifx \showDOI      \undefined \def \showDOI       #1{#1}\fi
\ifx \showISBNx    \undefined \def \showISBNx     #1{\unskip}     \fi
\ifx \showISBNxiii \undefined \def \showISBNxiii  #1{\unskip}     \fi
\ifx \showISSN     \undefined \def \showISSN      #1{\unskip}     \fi
\ifx \showLCCN     \undefined \def \showLCCN      #1{\unskip}     \fi
\ifx \shownote     \undefined \def \shownote      #1{#1}          \fi
\ifx \showarticletitle \undefined \def \showarticletitle #1{#1}   \fi
\ifx \showURL      \undefined \def \showURL       {\relax}        \fi
% invisible to TeX
\providecommand\bibfield[2]{#2}
\providecommand\bibinfo[2]{#2}
\providecommand\natexlab[1]{#1}
\providecommand\showeprint[2][]{arXiv:#2}

\bibitem[\protect\citeauthoryear{Baker, Bond, Corbett, Furman, Khorlin, Larson,
  Leon, Li, Lloyd, and Yushprakh}{Baker et~al\mbox{.}}{2011}]%
        {paper_baker_megastore}
\bibfield{author}{\bibinfo{person}{Jason Baker}, \bibinfo{person}{Chris Bond},
  \bibinfo{person}{James~C. Corbett}, \bibinfo{person}{JJ Furman},
  \bibinfo{person}{Andrey Khorlin}, \bibinfo{person}{James Larson},
  \bibinfo{person}{Jean-Michel Leon}, \bibinfo{person}{Yawei Li},
  \bibinfo{person}{Alexander Lloyd}, {and} \bibinfo{person}{Vadim Yushprakh}.}
  \bibinfo{year}{2011}\natexlab{}.
\newblock \showarticletitle{Megastore: Providing Scalable, Highly Available
  Storage for Interactive Services}. In \bibinfo{booktitle}{\emph{Proc. of
  CIDR}}.
\newblock


\bibitem[\protect\citeauthoryear{Bermbach}{Bermbach}{2014}]%
        {diss_bermbach}
\bibfield{author}{\bibinfo{person}{David Bermbach}.}
  \bibinfo{year}{2014}\natexlab{}.
\newblock \emph{\bibinfo{title}{Benchmarking Eventually Consistent Distributed
  Storage Systems}}.
\newblock \bibinfo{thesistype}{Ph.D. Dissertation}. \bibinfo{school}{Karlsruhe
  Institute of Technology}.
\newblock


\bibitem[\protect\citeauthoryear{Bermbach, Klems, Tai, and Menzel}{Bermbach
  et~al\mbox{.}}{2011}]%
        {paper_bermbach_metastorage}
\bibfield{author}{\bibinfo{person}{David Bermbach}, \bibinfo{person}{Markus
  Klems}, \bibinfo{person}{Stefan Tai}, {and} \bibinfo{person}{Michael
  Menzel}.} \bibinfo{year}{2011}\natexlab{}.
\newblock \showarticletitle{Metastorage: A federated cloud storage system to
  manage consistency-latency tradeoffs}. In \bibinfo{booktitle}{\emph{2011 IEEE
  4th International Conference on Cloud Computing}}. IEEE,
  \bibinfo{pages}{452--459}.
\newblock


\bibitem[\protect\citeauthoryear{Bermbach, Pallas, P\'{e}rez, Plebani,
  Anderson, Kat, and Tai}{Bermbach et~al\mbox{.}}{2017}]%
        {paper_bermbach_fog_vision}
\bibfield{author}{\bibinfo{person}{D. Bermbach}, \bibinfo{person}{F. Pallas},
  \bibinfo{person}{D.~Garc\'{i}a P\'{e}rez}, \bibinfo{person}{P. Plebani},
  \bibinfo{person}{M. Anderson}, \bibinfo{person}{R. Kat}, {and}
  \bibinfo{person}{S. Tai}.} \bibinfo{year}{2017}\natexlab{}.
\newblock \showarticletitle{A Research Perspective on Fog Computing}. In
  \bibinfo{booktitle}{\emph{Proc. of ISYCC}}. \bibinfo{publisher}{Springer}.
\newblock


\bibitem[\protect\citeauthoryear{Burrows}{Burrows}{2006}]%
        {paper_burrows_chubby}
\bibfield{author}{\bibinfo{person}{Mike Burrows}.}
  \bibinfo{year}{2006}\natexlab{}.
\newblock \showarticletitle{The Chubby Lock Service for Loosely-coupled
  Distributed Systems}. In \bibinfo{booktitle}{\emph{Proc. of OSDI}}.
  \bibinfo{publisher}{USENIX}.
\newblock


\bibitem[\protect\citeauthoryear{Chang, Dean, Ghemawat, Hsieh, Wallach,
  Burrows, Chandra, Fikes, and Gruber}{Chang et~al\mbox{.}}{2006}]%
        {paper_chang_bigtable}
\bibfield{author}{\bibinfo{person}{Fay Chang}, \bibinfo{person}{Jeffrey Dean},
  \bibinfo{person}{Sanjay Ghemawat}, \bibinfo{person}{Wilson~C. Hsieh},
  \bibinfo{person}{Deborah~A. Wallach}, \bibinfo{person}{Mike Burrows},
  \bibinfo{person}{Tushar Chandra}, \bibinfo{person}{Andrew Fikes}, {and}
  \bibinfo{person}{Robert~E. Gruber}.} \bibinfo{year}{2006}\natexlab{}.
\newblock \showarticletitle{Bigtable: A Distributed Storage System for
  Structured Data}. In \bibinfo{booktitle}{\emph{Proc. of OSDI}}.
  \bibinfo{publisher}{USENIX}.
\newblock


\bibitem[\protect\citeauthoryear{Chen, Chen, You, Ling, Liang, and
  Zimmermann}{Chen et~al\mbox{.}}{2016}]%
        {paper_chen_fog_video_surveillance}
\bibfield{author}{\bibinfo{person}{Ning Chen}, \bibinfo{person}{Yu Chen},
  \bibinfo{person}{Yang You}, \bibinfo{person}{Haibin Ling},
  \bibinfo{person}{Pengpeng Liang}, {and} \bibinfo{person}{Roger Zimmermann}.}
  \bibinfo{year}{2016}\natexlab{}.
\newblock \showarticletitle{Dynamic Urban Surveillance Video Stream Processing
  using Fog Computing}. In \bibinfo{booktitle}{\emph{Proc. of BigMM}}. IEEE.
\newblock


\bibitem[\protect\citeauthoryear{Confais, Lebre, and Parrein}{Confais
  et~al\mbox{.}}{2017a}]%
        {confais_object_2017}
\bibfield{author}{\bibinfo{person}{Bastien Confais}, \bibinfo{person}{Adrien
  Lebre}, {and} \bibinfo{person}{Benoit Parrein}.}
  \bibinfo{year}{2017}\natexlab{a}.
\newblock \showarticletitle{An Object Store Service for a Fog/Edge Computing
  Infrastructure Based on {IPFS} and a Scale-Out {NAS}}. In
  \bibinfo{booktitle}{\emph{Proc. of ICFEC}}. \bibinfo{publisher}{{IEEE}}.
\newblock


\bibitem[\protect\citeauthoryear{Confais, Lebre, and Parrein}{Confais
  et~al\mbox{.}}{2017b}]%
        {paper_confais_benchmarking_storage_for_fog}
\bibfield{author}{\bibinfo{person}{Bastien Confais}, \bibinfo{person}{Adrien
  Lebre}, {and} \bibinfo{person}{Beno{\^\i}t Parrein}.}
  \bibinfo{year}{2017}\natexlab{b}.
\newblock \showarticletitle{Performance Analysis of Object Store Systems in a
  Fog and Edge Computing Infrastructure}.
\newblock In \bibinfo{booktitle}{\emph{TLDKS}}. \bibinfo{publisher}{Springer}.
\newblock


\bibitem[\protect\citeauthoryear{Cooper, Ramakrishnan, Srivastava, Silberstein,
  Bohannon, Jacobsen, Puz, Weaver, and Yerneni}{Cooper et~al\mbox{.}}{2008}]%
        {paper_cooper_pnuts}
\bibfield{author}{\bibinfo{person}{Brian~F. Cooper}, \bibinfo{person}{Raghu
  Ramakrishnan}, \bibinfo{person}{Utkarsh Srivastava}, \bibinfo{person}{Adam
  Silberstein}, \bibinfo{person}{Philip Bohannon}, \bibinfo{person}{Hans-Arno
  Jacobsen}, \bibinfo{person}{Nick Puz}, \bibinfo{person}{Daniel Weaver}, {and}
  \bibinfo{person}{Ramana Yerneni}.} \bibinfo{year}{2008}\natexlab{}.
\newblock \showarticletitle{PNUTS: Yahoo!'s Hosted Data Serving Platform}.
\newblock \bibinfo{journal}{\emph{Proc. of VLDB}} (\bibinfo{year}{2008}).
\newblock


\bibitem[\protect\citeauthoryear{D\'{i}az, Mart\'{i}n, and Rubio}{D\'{i}az
  et~al\mbox{.}}{2016}]%
        {paper_diaz_challenges_iot_and_cloud}
\bibfield{author}{\bibinfo{person}{Manuel D\'{i}az}, \bibinfo{person}{Cristian
  Mart\'{i}n}, {and} \bibinfo{person}{Bartolom\'{e} Rubio}.}
  \bibinfo{year}{2016}\natexlab{}.
\newblock \showarticletitle{State-of-the-art, Challenges, and Open Issues in
  the Integration of Internet of Things and Cloud Computing}.
\newblock \bibinfo{journal}{\emph{Journal of Network and Computer
  Applications}} (\bibinfo{year}{2016}).
\newblock


\bibitem[\protect\citeauthoryear{Ghemawat, Gobioff, and Leung}{Ghemawat
  et~al\mbox{.}}{2003}]%
        {paper_ghemawat_google_file_system}
\bibfield{author}{\bibinfo{person}{Sanjay Ghemawat}, \bibinfo{person}{Howard
  Gobioff}, {and} \bibinfo{person}{Shun-Tak Leung}.}
  \bibinfo{year}{2003}\natexlab{}.
\newblock \showarticletitle{The Google File System}. In
  \bibinfo{booktitle}{\emph{Proc. of SOSP}}. \bibinfo{publisher}{ACM}.
\newblock


\bibitem[\protect\citeauthoryear{Grambow, Hasenburg, and Bermbach}{Grambow
  et~al\mbox{.}}{2018}]%
        {paper_grambow_fog_video_surveillance}
\bibfield{author}{\bibinfo{person}{Martin Grambow}, \bibinfo{person}{Jonathan
  Hasenburg}, {and} \bibinfo{person}{David Bermbach}.}
  \bibinfo{year}{2018}\natexlab{}.
\newblock \showarticletitle{Public Video Surveillance: Using the Fog to
  Increase Privacy}. In \bibinfo{booktitle}{\emph{Proc. of M4IoT}}.
  \bibinfo{publisher}{{ACM}}.
\newblock


\bibitem[\protect\citeauthoryear{Gupta and Ramachandran}{Gupta and
  Ramachandran}{2018}]%
        {gupta2018fogstore}
\bibfield{author}{\bibinfo{person}{Harshit Gupta} {and}
  \bibinfo{person}{Umakishore Ramachandran}.} \bibinfo{year}{2018}\natexlab{}.
\newblock \showarticletitle{Fogstore: A geo-distributed key-value store
  guaranteeing low latency for strongly consistent access}. In
  \bibinfo{booktitle}{\emph{Proceedings of the 12th ACM International
  Conference on Distributed and Event-based Systems}}. ACM,
  \bibinfo{pages}{148--159}.
\newblock


\bibitem[\protect\citeauthoryear{Gupta, Xu, and Ramachandran}{Gupta
  et~al\mbox{.}}{2018}]%
        {gupta2018datafog}
\bibfield{author}{\bibinfo{person}{Harshit Gupta}, \bibinfo{person}{Zhuangdi
  Xu}, {and} \bibinfo{person}{Umakishore Ramachandran}.}
  \bibinfo{year}{2018}\natexlab{}.
\newblock \showarticletitle{Datafog: Towards a holistic data management
  platform for the iot age at the network edge}. In
  \bibinfo{booktitle}{\emph{$\{$USENIX$\}$ Workshop on Hot Topics in Edge
  Computing (HotEdge 18)}}.
\newblock


\bibitem[\protect\citeauthoryear{Hasenburg, Grambow, and Bermbach}{Hasenburg
  et~al\mbox{.}}{2020}]%
        {paper_hasenburg_towards_fbase}
\bibfield{author}{\bibinfo{person}{Jonathan Hasenburg}, \bibinfo{person}{Martin
  Grambow}, {and} \bibinfo{person}{David Bermbach}.}
  \bibinfo{year}{2020}\natexlab{}.
\newblock \showarticletitle{{Towards A Replication Service for Data-Intensive
  Fog Applications}}. In \bibinfo{booktitle}{\emph{Proceedings of the 35th ACM
  Symposium on Applied Computing, Posters Track (SAC 2020)}}.
  \bibinfo{publisher}{ACM}.
\newblock


\bibitem[\protect\citeauthoryear{Hintjens}{Hintjens}{2013}]%
        {zeromq}
\bibfield{author}{\bibinfo{person}{Pieter Hintjens}.}
  \bibinfo{year}{2013}\natexlab{}.
\newblock \bibinfo{booktitle}{\emph{ZeroMQ: Messaging for Many Applications}}.
\newblock \bibinfo{publisher}{O'Reilly Media, Inc.}
\newblock


\bibitem[\protect\citeauthoryear{Hou, Li, Chen, Wu, Jin, and Chen}{Hou
  et~al\mbox{.}}{2016}]%
        {paper_hou_vehicular_computing}
\bibfield{author}{\bibinfo{person}{Xueshi Hou}, \bibinfo{person}{Yong Li},
  \bibinfo{person}{Min Chen}, \bibinfo{person}{Di Wu}, \bibinfo{person}{Depeng
  Jin}, {and} \bibinfo{person}{Sheng Chen}.} \bibinfo{year}{2016}\natexlab{}.
\newblock \showarticletitle{Vehicular Fog Computing: a Viewpoint of Vehicles as
  the Infrastructures}.
\newblock \bibinfo{journal}{\emph{IEEE TVT}} (\bibinfo{year}{2016}).
\newblock


\bibitem[\protect\citeauthoryear{Lin, Kemme, Patino-Martinez, and
  Jimenez-Peris}{Lin et~al\mbox{.}}{2007}]%
        {paper_lin_enhancing_edge_computing}
\bibfield{author}{\bibinfo{person}{Yi Lin}, \bibinfo{person}{Bettina Kemme},
  \bibinfo{person}{Marta Patino-Martinez}, {and} \bibinfo{person}{Ricardo
  Jimenez-Peris}.} \bibinfo{year}{2007}\natexlab{}.
\newblock \showarticletitle{Enhancing Edge Computing with Database
  Replication}. In \bibinfo{booktitle}{\emph{Proc. of SRDS}}. IEEE.
\newblock


\bibitem[\protect\citeauthoryear{Mayer, Gupta, Saurez, and Ramachandran}{Mayer
  et~al\mbox{.}}{2017}]%
        {paper_mayer_fogstore}
\bibfield{author}{\bibinfo{person}{Ruben Mayer}, \bibinfo{person}{Harshit
  Gupta}, \bibinfo{person}{Enrique Saurez}, {and} \bibinfo{person}{Umakishore
  Ramachandran}.} \bibinfo{year}{2017}\natexlab{}.
\newblock \showarticletitle{Fogstore: Toward a distributed data store for fog
  computing}. In \bibinfo{booktitle}{\emph{2017 IEEE Fog World Congress
  (FWC)}}. IEEE, \bibinfo{pages}{1--6}.
\newblock


\bibitem[\protect\citeauthoryear{Mortazavi, Salehe, Gomes, Phillips, and
  de~Lara}{Mortazavi et~al\mbox{.}}{2017}]%
        {mortazavi2017cloudpath}
\bibfield{author}{\bibinfo{person}{Seyed~Hossein Mortazavi},
  \bibinfo{person}{Mohammad Salehe}, \bibinfo{person}{Carolina~Simoes Gomes},
  \bibinfo{person}{Caleb Phillips}, {and} \bibinfo{person}{Eyal de Lara}.}
  \bibinfo{year}{2017}\natexlab{}.
\newblock \showarticletitle{Cloudpath: a multi-tier cloud computing framework}.
  In \bibinfo{booktitle}{\emph{Proceedings of the Second ACM/IEEE Symposium on
  Edge Computing}}. ACM, \bibinfo{pages}{20}.
\newblock


\bibitem[\protect\citeauthoryear{Naas, Parvedy, Boukhobza, and Lemarchand}{Naas
  et~al\mbox{.}}{2017}]%
        {naas2017ifogstor}
\bibfield{author}{\bibinfo{person}{Mohammed~Islam Naas},
  \bibinfo{person}{Philippe~Raipin Parvedy}, \bibinfo{person}{Jalil Boukhobza},
  {and} \bibinfo{person}{Laurent Lemarchand}.} \bibinfo{year}{2017}\natexlab{}.
\newblock \showarticletitle{iFogStor: an IoT data placement strategy for fog
  infrastructure}. In \bibinfo{booktitle}{\emph{2017 IEEE 1st International
  Conference on Fog and Edge Computing (ICFEC)}}. IEEE,
  \bibinfo{pages}{97--104}.
\newblock


\bibitem[\protect\citeauthoryear{Plebani, Garcia-Perez, Anderson, Bermbach,
  Cappiello, Kat, Pallas, Pernici, Tai, and Vitali}{Plebani
  et~al\mbox{.}}{2017}]%
        {paper_plebani_ditas_vision}
\bibfield{author}{\bibinfo{person}{Pierluigi Plebani}, \bibinfo{person}{David
  Garcia-Perez}, \bibinfo{person}{Maya Anderson}, \bibinfo{person}{David
  Bermbach}, \bibinfo{person}{Cinzia Cappiello}, \bibinfo{person}{Ronen~I Kat},
  \bibinfo{person}{Frank Pallas}, \bibinfo{person}{Barbara Pernici},
  \bibinfo{person}{Stefan Tai}, {and} \bibinfo{person}{Monica Vitali}.}
  \bibinfo{year}{2017}\natexlab{}.
\newblock \showarticletitle{Information logistics and fog computing: The DITAS
  approach}. In \bibinfo{booktitle}{\emph{CEUR Workshop Proceedings}}. CEUR-WS.
\newblock


\bibitem[\protect\citeauthoryear{Pradhan, Dubey, Khare, Nannapaneni, Gokhale,
  Mahadevan, Schmidt, and Lehofer}{Pradhan et~al\mbox{.}}{2017}]%
        {paper_pradhan_chariot_edge_iot}
\bibfield{author}{\bibinfo{person}{Subhav Pradhan}, \bibinfo{person}{Abhishek
  Dubey}, \bibinfo{person}{Shweta Khare}, \bibinfo{person}{Saideep
  Nannapaneni}, \bibinfo{person}{Aniruddha Gokhale}, \bibinfo{person}{Sankaran
  Mahadevan}, \bibinfo{person}{Douglas~C Schmidt}, {and}
  \bibinfo{person}{Martin Lehofer}.} \bibinfo{year}{2017}\natexlab{}.
\newblock \showarticletitle{Chariot: Goal-driven Orchestration Middleware for
  Resilient IoT Systems}.
\newblock \bibinfo{journal}{\emph{ACM TCPS}} (\bibinfo{year}{2017}).
\newblock


\bibitem[\protect\citeauthoryear{Psaras, Ascigil, Rene, Pavlou, Afanasyev, and
  Zhang}{Psaras et~al\mbox{.}}{2018}]%
        {psaras2018mobile}
\bibfield{author}{\bibinfo{person}{Ioannis Psaras}, \bibinfo{person}{Onur
  Ascigil}, \bibinfo{person}{Sergi Rene}, \bibinfo{person}{George Pavlou},
  \bibinfo{person}{Alex Afanasyev}, {and} \bibinfo{person}{Lixia Zhang}.}
  \bibinfo{year}{2018}\natexlab{}.
\newblock \showarticletitle{Mobile data repositories at the edge}. In
  \bibinfo{booktitle}{\emph{$\{$USENIX$\}$ Workshop on Hot Topics in Edge
  Computing (HotEdge 18)}}.
\newblock


\bibitem[\protect\citeauthoryear{Ryden, Oh, Chandra, and Weissman}{Ryden
  et~al\mbox{.}}{2014}]%
        {paper_ryden_nebula}
\bibfield{author}{\bibinfo{person}{Mathew Ryden}, \bibinfo{person}{Kwangsung
  Oh}, \bibinfo{person}{Abhishek Chandra}, {and} \bibinfo{person}{Jon
  Weissman}.} \bibinfo{year}{2014}\natexlab{}.
\newblock \showarticletitle{Nebula: Distributed Edge Cloud for Data Intensive
  Computing}. In \bibinfo{booktitle}{\emph{Proc. of IC2E}}. IEEE.
\newblock


\bibitem[\protect\citeauthoryear{Shi, Cao, Zhang, Li, and Xu}{Shi
  et~al\mbox{.}}{2016}]%
        {paper_shi_edge_challenges}
\bibfield{author}{\bibinfo{person}{W. Shi}, \bibinfo{person}{J. Cao},
  \bibinfo{person}{Q. Zhang}, \bibinfo{person}{Y. Li}, {and}
  \bibinfo{person}{L. Xu}.} \bibinfo{year}{2016}\natexlab{}.
\newblock \showarticletitle{Edge Computing: Vision and Challenges}.
\newblock \bibinfo{journal}{\emph{IEEE IoT-J}} (\bibinfo{year}{2016}).
\newblock


\bibitem[\protect\citeauthoryear{Shi and Dustdar}{Shi and Dustdar}{2016}]%
        {paper_shi_fog_computing_definition}
\bibfield{author}{\bibinfo{person}{W. Shi} {and} \bibinfo{person}{S. Dustdar}.}
  \bibinfo{year}{2016}\natexlab{}.
\newblock \showarticletitle{The Promise of Edge Computing}.
\newblock \bibinfo{journal}{\emph{Computer}} (\bibinfo{year}{2016}).
\newblock


\bibitem[\protect\citeauthoryear{Shneidman, Pietzuch, Ledlie, Roussopoulos,
  Seltzer, and Welsh}{Shneidman et~al\mbox{.}}{2004}]%
        {hourglass}
\bibfield{author}{\bibinfo{person}{Jeff Shneidman}, \bibinfo{person}{Peter
  Pietzuch}, \bibinfo{person}{Jonathan Ledlie}, \bibinfo{person}{Mema
  Roussopoulos}, \bibinfo{person}{Margo Seltzer}, {and} \bibinfo{person}{Matt
  Welsh}.} \bibinfo{year}{2004}\natexlab{}.
\newblock \showarticletitle{Hourglass: An Infrastructure for Connecting Sensor
  Networks and Applications}. In \bibinfo{booktitle}{\emph{Techn. Rep. TR-21-04
  Harvard University}}.
\newblock


\bibitem[\protect\citeauthoryear{Vogels}{Vogels}{2008}]%
        {paper_vogels_eventually_consistent}
\bibfield{author}{\bibinfo{person}{Werner Vogels}.}
  \bibinfo{year}{2008}\natexlab{}.
\newblock \showarticletitle{Eventually Consistent}.
\newblock \bibinfo{journal}{\emph{ACM Queue}} (\bibinfo{year}{2008}).
\newblock


\bibitem[\protect\citeauthoryear{Zhang, Mor, Kolb, Chan, Lutz, Allman,
  Wawrzynek, Lee, and Kubiatowicz}{Zhang et~al\mbox{.}}{2015}]%
        {paper_zhang_cloud_is_not_enough_GDP}
\bibfield{author}{\bibinfo{person}{Ben Zhang}, \bibinfo{person}{Nitesh Mor},
  \bibinfo{person}{John Kolb}, \bibinfo{person}{Douglas~S Chan},
  \bibinfo{person}{Ken Lutz}, \bibinfo{person}{Eric Allman},
  \bibinfo{person}{John Wawrzynek}, \bibinfo{person}{Edward~A Lee}, {and}
  \bibinfo{person}{John Kubiatowicz}.} \bibinfo{year}{2015}\natexlab{}.
\newblock \showarticletitle{The Cloud is Not Enough: Saving IoT from the
  Cloud.}. In \bibinfo{booktitle}{\emph{HotStorage}}.
  \bibinfo{publisher}{USENIX}.
\newblock


\end{thebibliography}

\end{document}